\documentclass[
reprint,
superscriptaddress,
nofootinbib,
amsmath,amssymb,
aps,
pra,
noeprint,
]{revtex4-2}

\usepackage{graphicx}
\usepackage[english]{babel}
\usepackage{appendix}
\usepackage[ruled]{algorithm2e}
\usepackage{dcolumn}
\usepackage{bm}
\usepackage{color}
\usepackage{booktabs,eqparbox}
\usepackage{braket}
\usepackage{dsfont}
\usepackage{placeins}
\usepackage{paralist}
\usepackage{comment}
\usepackage{multirow}
\usepackage{mhchem}
\usepackage[utf8]{inputenc}
\usepackage{pgfplots}
\usepackage{amsmath}
\usepackage{subcaption}
\usepackage{ragged2e}
\usepackage{graphicx, import}
\usepackage{tikz}
\usepgflibrary{plotmarks}
\usepgflibrary[plotmarks]
\usetikzlibrary{plotmarks, positioning}
\usetikzlibrary[plotmarks]
\usepackage{csquotes}
\usetikzlibrary{shapes,arrows,fit, backgrounds, quantikz2}
\usepackage{yquant, quantikz}
\useyquantlanguage{groups}
\usetikzlibrary{calc}

\usepackage{textgreek}
\usepackage{siunitx}
\usepackage[nolist,nohyperlinks]{acronym}
\usepackage[colorlinks=true, allcolors=magenta]{hyperref}
\usepackage{cleveref}

\usepgfplotslibrary{groupplots,dateplot}
\usetikzlibrary{patterns,shapes.arrows}
\pgfplotsset{compat=newest}

\makeatletter
\let\oldtheequation\theequation
\renewcommand\tagform@[1]{\maketag@@@{\ignorespaces#1\unskip\@@italiccorr}}
\renewcommand\theequation{(\oldtheequation)}
\makeatother

\definecolor{Lochinvar}{RGB}{49, 160, 143}
\definecolor{hblue}{RGB}{86, 175, 255}

\newcommand{\imag}{\mathrm{i}}
\newcommand{\lb}{\left[}
\newcommand{\rb}{\right]}
\newcommand{\lp}{\left(}
\newcommand{\rp}{\right)}

\newcommand{\ra}{\rightarrow}

\newcommand{\Hbg}{\ensuremath{H_Z}}
\newcommand{\initial}{\ensuremath{\ket{\mathrm{HF}}}}
\newcommand{\brainitial}{\ensuremath{\bra{\mathrm{HF}}}}
\newcommand{\tr}{\mathrm{tr}}

\addto\extrasenglish{%
}

\addto\extrasenglish{%
}

\graphicspath{{Figures/}}

\begin{document}

\title{Ground State Energy via Adiabatic Evolution and Phase Measurement for a Molecular Hamiltonian on an Ion-Trap Quantum Computer}

\newcommand{\QTNGER}{Quantinuum, Leopoldstrasse 180, 80804 Munich, Germany}
\newcommand{\ERL}{Department of Physics, Friedrich-Alexander Universit\"at Erlangen-N\"urnberg, Erlangen, Germany}
\newcommand{\MPL}{Max Planck Institute for the Science of Light, Staudtstraße 2, 91058 Erlangen, Germany}

\author{Ludwig N\"utzel}
\affiliation{\ERL}
\affiliation{\QTNGER}

\author{Michael J. Hartmann}
\affiliation{\ERL}
\affiliation{\MPL}

\author{Henrik Dreyer}
\affiliation{\QTNGER}

\author{Etienne Granet}
\affiliation{\QTNGER}

    \begin{abstract}
       {
        Estimating molecular ground-state energies is a central application of quantum computing, requiring both the preparation of accurate quantum states and efficient energy readout. Understanding the effect of hardware noise on these experiments is crucial to distinguish errors that have low impact, errors that can be mitigated, and errors that must be reduced at the hardware level.
We ran a state preparation and energy measurement protocol on an ion-trap quantum computer, without any non-scalable off-loading of computational tasks to classical computers, and  show that leakage errors are the main obstacle to chemical accuracy.
More specifically, we apply adiabatic state preparation to prepare the ground state of a six-qubit encoding of the \ce{H3+} molecule and extract its energy using a noise-resilient variant of iterative quantum phase estimation. 
Our results improve upon the classical Hartree–Fock energy.
Analyzing the effect of hardware noise on the result, we find that while coherent and incoherent noise have little influence, the hardware results are mainly impacted by leakage errors.
Absent leakage errors, noisy numerical simulations show that with our experimental settings we would have achieved close to chemical accuracy, even shot noise included.
These insights highlight the importance of targeting leakage suppression in future algorithm and hardware development.
       }
    \end{abstract}

\maketitle

\section{Introduction}
Accurately estimating molecular \acp{GSE} is one of the central challenges in quantum chemistry and a key application area for quantum computing~\cite{cao2019quantum,clinton2024towards,bauer2020quantum}.
Classical methods such as Hartree-Fock, density functional theory or coupled cluster methods provide useful approximations but fail to capture essential electron correlation effects, limiting their predictive power for strongly correlated systems~\cite{Kohn1965,cizek1966}.
In these cases, more sophisticated multi-reference methods such as complete active space configuration interaction or \ac{CASSCF} are needed for an appropriate description of electronic correlations~\cite{Roca2012}.
However, these classical methods scale exponentially with system size and eventually become prohibitively expensive.

Quantum algorithms are expected to deliver systematic improvements in this scenario.
However, their practical realization is hindered by the difficulty of preparing accurate quantum states, estimating observables with high precision, and mitigating hardware noise. These obstacles are considerably increased by the high precision on \acp{GSE} required in chemistry applications.
Among the available approaches, \ac{ASP} is a theoretically attractive means to prepare suitable states because it avoids parameter optimization, the notorious barren plateau phenomenon, and the need for repeated intermediate observable estimation~\cite{Granet2024,McClean2018,Holmes2022,Cerezo2021}.
\begin{figure}[ht!]
    \centering
    \includegraphics[width=\linewidth]{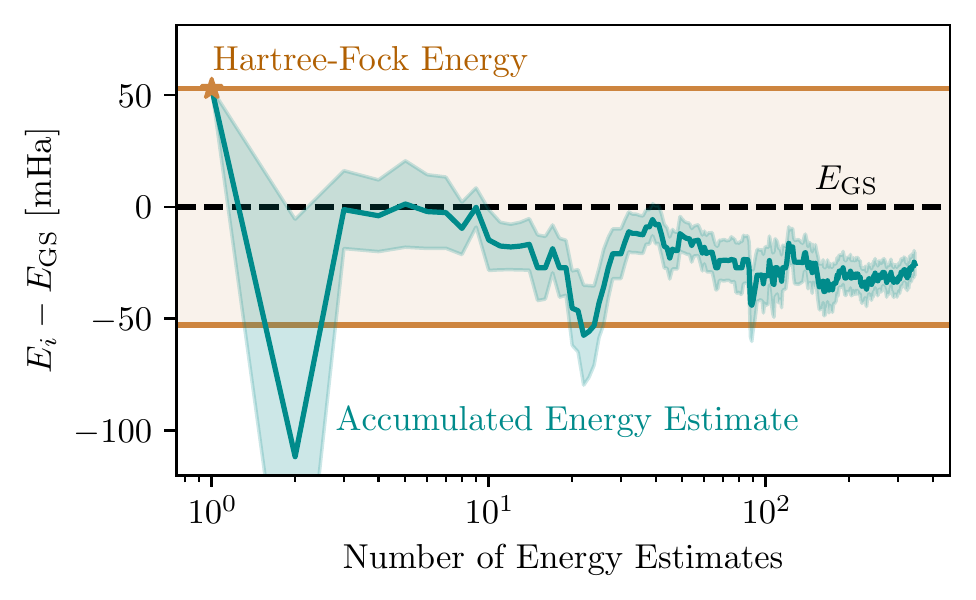}
    \caption{\label{fig:energy_convergence_variance}\textbf{Hardware result: Accumulated energy estimates approaching the ground-state energy $E_{\mathrm{GS}}$.}
    The solid teal curve shows hardware results after post-selection on parity and projection onto the Hartree-Fock state; the shaded region indicates statistical uncertainty.
    The horizontal dashed line marks $E_{\mathrm{GS}}$.
    The orange line represents the Hartree-Fock energy, illustrating the deviation of the classical estimate from $E_{\mathrm{GS}}$.
    In an ideal noiseless scenario, $E_i$, the accumulated energy estimate using all measurements up to the $i$-th, would converge to $E_{\mathrm{GS}}$ as $i$ increases.
    }
\end{figure}
Its actual runtime and speed of convergence are however difficult to evaluate beforehand and are mostly determined from heuristics or extrapolated from simulable system sizes.  Moreover, it requires the implementation of dynamics generated by time-dependent molecular Hamiltonians, which is a difficult task.
Performing \ac{ASP} on real quantum hardware has long been considered challenging due to these requirements, though several recent works have demonstrated its feasibility on small‑scale devices~\cite{du2010nmr,chen2020demonstration,hegade2021shortcuts,francis2022determining,cugini2024spectral,burton2025excited,karacan2025filter}.

Another key bottleneck is the \ac{GSE} estimation.
Using state-of-the-art measurement strategies, directly sampling the expectation value of a molecular ground state to chemical precision requires millions of shots, for any system size~\cite{gresch2025guaranteed}.
Other recent approaches combine computational basis measurements and quantum subspace diagonalization to circumvent this problem, but the scalability of this approach still suffers from sampling challenges~\cite{kanno2023quantum,robledo2025chemistry,nutzel2025solving,reinholdt2025critical}.
An algorithm that does not have a sampling overhead is \ac{QPE}, which however comes at the cost of having to implement expensive controlled-$U^{2^k}$ operations, where $U$ is usually the operator exponential of the target Hamiltonian. Chemical precision requires $2^k\sim 10^3$, which is prohibitively expensive.
Experimental demonstrations of \ac{QPE}, or variants thereof, have thus been limited to energy estimations of trivial or easy-to-prepare ground states at small system sizes~\cite{mohammadbagherpoor2019improved,liu2023full,yamamoto2306demonstrating}.
Moreover, experiments that combine \ac{ASP} and \ac{QPE} have been limited to one-qubit systems~\cite{du2010nmr}.

In this work, we demonstrate \ac{ASP} and \ac{GSE} estimation in a more practical fashion on a trapped-ion quantum computer. By practical we mean that we only make use of purely quantum, \textit{ab initio} methods, and we do not rely on off-loading a non-scalable parts of the quantum algorithm to classical computers.
Under the assumption that adiabatic times scale mildly with system size, we therefore expect this algorithm to be applicable at large system sizes.
We make use of recent theoretical advances that allow for an implementation of \ac{ASP} that is free of Trotter errors~\cite{Granet2023}, and employ a variant of iterative quantum phase estimation to efficiently read out the energy~\cite{Granet2024}.
The system chosen is a six-qubit encoding of \ce{H3+}.
After applying classical post-processing techniques, the resulting expectation value on hardware has a deviation of $-25.5\pm 5.3$\,mHa from the exact \ac{GSE}.
With the classical \ac{HF} estimate deviating $52.8$\,mHa from the exact \ac{GSE}, the estimate we obtain from quantum hardware is closer to the desired value by more than 5 standard deviations.

Nevertheless, the accuracy obtained on quantum hardware is lower than what was expected from simple noise models with realistic noise rates. To evaluate the impact of different noise sources, we analyze coherent, incoherent and leakage noise models in classical simulations.
The central outcome of this analysis is the identification of leakage errors as the most likely dominant hardware limitation.
Classical noisy simulations that only include coherent and incoherent channels predict energy estimates close to chemical accuracy, even when shot noise is included.
The experimental deviation from these predictions is reproduced only when leakage is taken into account.
We conclude that while the algorithm of Ref.~\cite{Granet2024} is robust against coherent and incoherent noise, there remains a strong sensitivity to leakage errors.
These findings underscore the critical need to address leakage suppression in the advancement of future algorithms and hardware development.

\section{Methods\label{sec:methods}}
\begin{figure*}[ht!]
    \centering
    \input{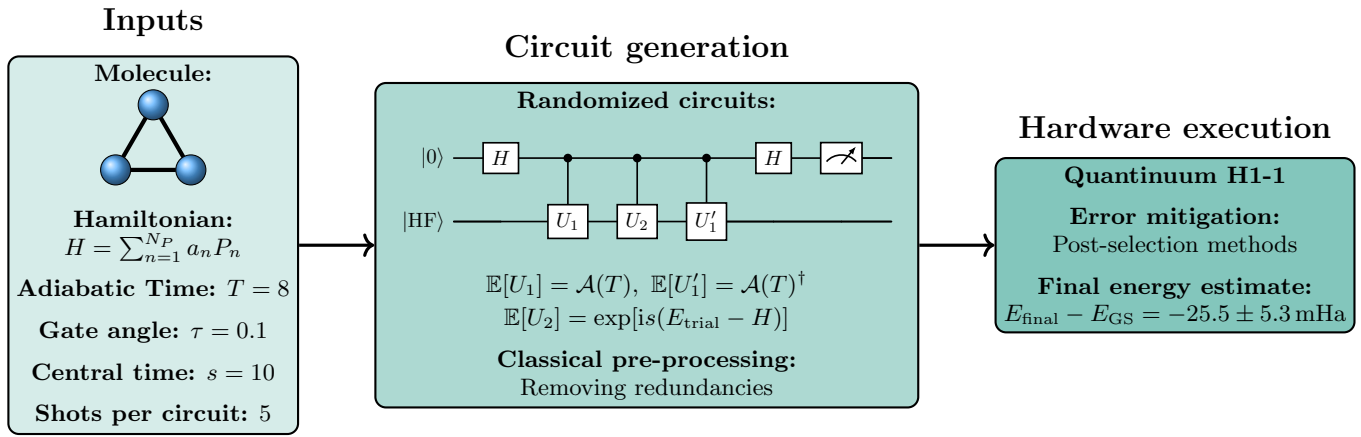}
    \caption{\label{fig:workflow}\textbf{Workflow diagram.}
    \textit{Left: Algorithm inputs.} The Pauli Hamiltonian $H$ is derived from the chosen molecular geometry, and a suitable adiabatic time is picked.
    The choice of gate angle $\tau$ and central time $s$ is governed e.g. by a maximum allowed circuit depth.
    The number of single-shot measurements is passed as an additional constraint.
    \textit{Center: Circuit generation.} The given inputs allow the generation of a certain number of circuits. Redundancies in circuits can be removed in classical pre-processing steps.
    \textit{Right: Hardware execution.} The circuits are executed on Quantinuum's H1-1 trapped-ion quantum computer.
    The noisy results can be post-selected, enhancing the quality of the final energy estimate.
    }
\end{figure*}

This section details the individual parts contributing to the experiment.
A high-level overview of the workflow is shown in \autoref{fig:workflow}.
\subsection{Problem Structure}
We aim at estimating the ground state energy of an electronic structure Hamiltonian,
\begin{align}
    H_\mathrm{el.} = \sum_{ij,\sigma} t_{ij}^\sigma c_{i\sigma}^\dagger c_{j\sigma}
    + \frac{1}{2}\sum_{ijkl, \sigma\sigma^\prime} t_{ijkl}^{\sigma\sigma^\prime} c_{i \sigma}^\dagger c_{j\sigma^\prime}^\dagger c_{k\sigma}c_{l\sigma^\prime},
    \label{eq:Hamiltonian_el}
\end{align}
where $c^\dagger$ and $c$ are fermionic creation and annihilation operators, $\{i,j,k,l\}$ index spatial orbitals, $\sigma \in \{\uparrow, \downarrow\}$ denotes the spin, and $t_{ij}$ and $t_{ijkl}$ are the real-valued one- and two-body integrals in the basis that is given by the molecular orbitals obtained from a self-consistent field calculation.
The fermionic ladder operators obey the canonical anti-commutation relations $\{c_{i,\sigma}^\dagger,c_{j,\sigma^\prime}\} = \delta_{ij}\delta_{\sigma \sigma^\prime}$ and $\{c_{i,\sigma},c_{j,\sigma^\prime}\} = 0$.
To implement them on a quantum computer, we employ the \ac{JW} encoding to map the fermionic operators to Pauli operators $\{I, X, Y, Z\}$,
\begin{align}
    c_m = \frac{1}{2}(X_m + \imag Y_m) Z_{m-1} \cdots Z_1,
    \label{eq:JW_mapping}
\end{align}
where the spatial orbital and spin indices are combined to a single-qubit index $m$.
The resulting $L$-qubit Hamiltonian can be expressed as a sum of Paulis $P_n\in\{I,X,Y,Z\}^{\otimes L}$,
\begin{align}
    H = \sum_{n=1}^{N_P} a_n P_n
\end{align}
where $N_P$ is the number of Pauli terms and $a_n$ are real coefficients.

\subsection{Adiabatic State Preparation\label{sec:ASP}}
To determine the ground state energy of $H$, we first need to prepare an appropriate quantum state.
In this work we prepare the ground state of $H$ by adiabatically evolving the ground state of a simple starting Hamiltonian \Hbg{} to the ground state of the target Hamiltonian $H$.
We thus split the Hamiltonian into two parts, $H = \Hbg + H_I$, and define the time-dependent Hamiltonian
\begin{align}
    H(u) = \Hbg + w(u)H_I,
\end{align}
where $\Hbg = \sum_{n=N_I+1}^{N_P} a_n P_n$ contains only single-$Z$ Pauli operators, $H_I = \sum_{n=1}^{N_I} a_n P_n$ contains all remaining terms, and $N_I$ is the number of terms in $H_I$.
We further define the \textit{interaction norm} of the interaction Hamiltonian $H_I$ as
\begin{align}
    \mu_I = \sum_{n=1}^{N_I} |a_n|,\label{eq:mu_I}
\end{align}
on which the runtime of our algorithm will mainly depend.
With the sweep function fulfilling $w(0)=0,\,w(1)=1$, the ground state of $H$ is obtained by adiabatically evolving the ground state \initial{} of $H(0)=\Hbg$ to that of $H(1)=\Hbg + H_I$. The crucial assumption here is that these two states are adiabatically well connected via the sweep function $w$.
The time-ordered operator that implements the \ac{ASP} is
\begin{align}
    \mathcal{A}(T) = \mathcal{T} \exp\lp \imag\int_0^T H\lp \frac{t}{T} \rp \mathrm{d}t \rp,
\end{align}
where $T$ is the duration of the adiabatic sweep, and larger $T$ yields a state closer to the exact ground state of $H$.\par%
A standard way of implementing $\mathcal{A}(T)$ on a quantum computer is to decompose it using a Trotter-Suzuki approximation.
Here one first needs to discretize the time evolution into a finite number of time steps, and then each resulting operator exponential needs to be decomposed\ \cite{sun2020trotterized,suzuki1985decomposition}.
These approximation errors lead to larger contributions of excited states which can quickly become a problem in cases where very high precision is needed, as is the case in quantum chemistry \cite{Granet2024}.
To avoid this, we employ an alternative method without discretization error and compare both approaches in\ \autoref{app:trotter}.\par%
The \textit{TETRIS} (time evolution through random independent sampling) algorithm provides a randomization procedure that enables exact computation of time-dependent unitary observables on a digital quantum computer without discretization error\ \cite{Granet2023}.
We outline the algorithm in Appendix~\ref{app:tetris} and refer to Refs.\ \cite{Granet2023, Granet2024} for further details.\par%
The main result of the algorithm is that it produces randomized unitaries that on average exactly reproduce a time evolution operator, up to some rescaling factor.
Denoting by $U$ the random unitary generated by the algorithm, its statistical average $\mathbb{E}$ is
\begin{align}
    \mathbb{E}[U] = \mathrm{e}^{-\tan(\tau/2)\zeta T \mu_I} \mathcal{A}(T),
\end{align}
where $0<\tau<\pi/2$ is a chosen gate angle, and $\zeta$ is a constant of order $\mathcal{O}(1)$.
The adiabatic state preparation is therefore implemented exactly up to an attenuation factor $\lambda\coloneq \mathrm{e}^{-\tan(\tau/2)\zeta T \mu_I}$.
Importantly, this is achieved with a finite average number of \acp{TQG} per circuit,
\begin{align}
    N_\mathrm{TQG} = \frac{\zeta T \mu_I g}{\sin\tau},\quad g=\frac{\sum_{n=1}^{N_I}|a_n| g_n}{\sum_{n=1}^{N_I}|a_n|},
\end{align}
where $g_n$ is the number of \acp{TQG} required to implement $\mathrm{e}^{\imag \tau P_n}$\ \cite{Granet2024}.
Since $\mu_I$, cf. \autoref{eq:mu_I}, dampens the attenuation factor and increases the \ac{TQG} count, it is the main factor limiting the efficiency of this algorithm.
Strategies for reducing $\mu_I$ will thus be discussed later in \autoref{sec:reducing_norm}.

\subsection{Energy Estimation\label{subsec:energy_estimation}}
Once the ground state
\begin{align}
    \ket{\psi(T)} = \mathcal{A}(T)\initial
\end{align} is prepared on a quantum computer, our goal is to determine its energy $E_\mathrm{GS}$.
Promising but still costly approaches are \ac{QPE} or \ac{iQPE}. In this work we employ a variant of phase estimation that was introduced in Ref.\ \cite{Granet2024}, which we briefly summarize here, and which has a lower gate count than \ac{QPE} but larger shot requirement.\par%
The core of the algorithm relies on measuring
\begin{align}
    \rho(s) = \mathfrak{Im}\bra{\psi(T)}\mathrm{e}^{\imag s (E_\mathrm{trial} - H)}\ket{\psi(T)},\label{eq:rho_definition}
\end{align}
where $E_\mathrm{trial}$ is a trial energy, $\mathfrak{Im}(x)$ denotes the imaginary part of $x$, and $s$ is the central time \cite{Granet2024}, constrained by $0 < s < \pi / |\delta|$ with $\delta = E_\mathrm{trial} - E_\mathrm{GS}$.
Choosing two trial energies
\begin{align}
    E_\mathrm{trial}^\pm = E_\mathrm{guess} \pm \epsilon,
\end{align}
where $E_\mathrm{guess}$ is a current estimate of the \ac{GSE} and $\epsilon$ is a manually chosen parameter, let $\rho_\pm$ denote $\rho(s)$ estimated at fixed $s$ and $E_\mathrm{trial}^\pm$.
The \ac{GSE} can then be estimated as\ \cite{Granet2024}
\begin{align}
    E_\mathrm{GSE} = E_\mathrm{guess} + \frac{1}{s}\arctan\lp \tan(s\epsilon)\frac{\rho_+ + \rho_-}{\rho_- - \rho_+} \rp.\label{eq:GSE_estimate}
\end{align}
The benefit of this method is that it leverages information about the amplitude of $\rho(s)$, rather than just its sign.
Further, if the main source of error is depolarizing noise, the values of $\rho_\pm$ will be damped similarly, so the noise largely cancels when computing the ratio\ \cite{Granet2024}.
We will later test the robustness of this estimator against noise numerically.\par%
The quantity $\rho(s)$ can be obtained via a Hadamard test.
In this setup, the initial state $\initial$ is evolved via the unitaries $U_1,U_1^\prime$ and $U_2$, where $U_1$ and $U_1^\prime$ are randomized implementations of $\mathcal{A}(T)$ and $\mathcal{A}(T)^\dagger$, and $U_2$ on average implements $\mathrm{e}^{\imag s (E_\mathrm{trial} - H)}$.
The resulting circuits, controlled by an ancilla qubit, prepare states of the form $U_1^\prime U_2 U_1 \initial$.
The statistical average over the ensemble of such circuits then is
\begin{align}
    \rho(s) = \lp \lambda_\mathcal{A}^2 \lambda_s \rp^{-1} \mathbb{E}\lb\mathfrak{Im} \brainitial U_1^\prime U_2 U_1\initial\rb,
\end{align}
where $\lambda_\mathcal{A}$ and $\lambda_s$ are the attenuation factors from the implementations of $U_1^{(\prime)}$ and $U_2$, respectively.

\subsection{Reducing the norm\label{sec:reducing_norm}}
As discussed in \autoref{sec:ASP}, the efficiency of the randomization algorithm is primarily limited by the 1-norm $\mu_I$ of the interaction Hamiltonian.
Reducing $\mu_I$ is therefore crucial for improving performance, and we will discuss a way of doing so in this section.\par%
Let $\mathcal{S}$ be a set of conserved quantities satisfying
\begin{align}
    [H(u), \xi_i] = 0\quad \forall\, u\in[0,1],\,\forall\, \xi_i \in \mathcal{S}.
\end{align}
In the subspace of all states $\ket{\phi}$ with $\xi_i\ket{\phi} = \overline{\xi}_i\ket{\phi}$, $\overline{\xi}_i\in\mathbb{R}$, the spectrum of $H(u)$ is invariant under the transformation
\begin{align}
    H(u)\ \ra \ H(u) + \sum_{i=1}^{|\mathcal{S}|}\alpha_i\lp\xi_i^{m_i} - \overline{\xi}_i^{m_i}\rp,
\end{align}
where $\alpha_i\in\mathbb{R}$ and $m_i\geq 0$ are parameters chosen to minimize $\mu_I$.
For an electronic structure Hamiltonian as in \autoref{eq:Hamiltonian_el}, the particle numbers in each spin sector, $n_\sigma = \sum_i c_{i\sigma}^\dagger c_{i\sigma}$, and the total particle number $n_\mathrm{tot}=n_\uparrow + n_\downarrow$ are conserved. Note that since single-$Z$ terms do not contribute to $\mu_I$, setting $m_i = 1$ has no effect.
We therefore choose $m_i = 2$ and define the parameterized Hamiltonian
\begin{align}
\begin{aligned}
    H(u, \{\alpha_i\}) = H(u) &+ \alpha_0 \lp n_\uparrow^2 - \overline{n}_\uparrow^2 \rp + \alpha_1 \lp n_\downarrow^2 - \overline{n}_\downarrow^2 \rp\\
    &+ \alpha_2 \lp n_\mathrm{tot}^2 - \overline{n}_\mathrm{tot}^2 \rp.
\end{aligned}
\label{eq:param_hamiltonian}
\end{align}
The squared number operators give rise to double-$Z$ Pauli operators, which also appear in the Hamiltonian due to density-density interactions.
Optimizing the parameters $\{\alpha_i\}$ reduces their contribution to $\mu_I$.

\subsection{Further improvements\label{subsec:further_improvements}}
Once a randomized circuit has been generated, further optimizations can be applied via classical pre- and post-processing and we make use of these in our study.
\paragraph{Parity-based operator reduction.}
All circuits are products of exponentials of Pauli operators $\exp(\imag \theta_i P_i)$, where each $P_i$ appears in the Hamiltonian.
It follows from \autoref{eq:Hamiltonian_el} and \ref{eq:JW_mapping} that every $P_i$ commutes with the parity operator $\Pi = Z_L Z_{L-1} \dots Z_1$, so parity is conserved throughout the circuit.
If the initial state $\initial$ (the ground state of $\Hbg$) satisfies $\Pi \initial = \eta \initial$ with $\eta = \pm 1$, then any intermediate state $\ket{\phi}$ also satisfies $\Pi \ket{\phi} = \eta \ket{\phi}$.
This allows us to rewrite any operator
\begin{align}
    \exp(\imag \theta_i P_i)\ \ra \exp(\imag \eta \theta_i P_i \Pi)
\end{align}
The point is that we can choose to only do this transform for operators where the number of Pauli-$Z$s is larger than the number of identities $I$.
In the most extreme case, an $L$-qubit operator can be reduced to a four-qubit operator, significantly lowering circuit depth:
\begin{align}
    \exp(\imag\theta_i XZ\cdots ZXYY)\ \ra \ \exp(\imag\eta\theta_i  YI\cdots IYXX).
\end{align}
\paragraph{Occupation-number-based operator reduction.}
The initial state $\initial$ is a single computational basis state, so the occupation number of each qubit $q$ is fixed: $\overline{n}_q = \brainitial n_q\initial \in \{0,1\}$.
Let $\mathcal{W}$ be the set of qubits with fixed occupations, $C = \prod_i \exp(\imag \theta_i P_i)$ the circuit, and $P_i^{(q)}$ the single-qubit Pauli acting on qubit $q$.
We can then apply the procedure in \autoref{alg:removing_Z}.
\begin{algorithm}
\caption{Occupation-number-based operator reduction\label{alg:removing_Z}}
\For{$\exp(\imag \theta_i P_i)$ in $C$}{
  \eIf{$|\mathcal{W}|>0$}{
    \ForEach{$q$ in $\mathcal{W}$}{
        \uIf{$P_i^{(q)}$ is $X$ or $Y$}{
            $\mathcal{W} \gets \mathcal{W}\setminus q$\;
        }
        \uElseIf{$P_i^{(q)}$ is $Z$}{
            $P_i^{(q)}\gets (-1)^{\overline{n}_q}I$\;
        }
    }
  }{
    break;
  }
}
\end{algorithm}
This algorithm effectively replaces $Z$s with $(-1)^{\overline{n}_q}I$ wherever possible, shortening the first parts of the circuit.\par%
\paragraph{Alternating sign of time evolution.}
\autoref{eq:rho_definition} defines $\rho$ with the operators $\mathcal{A}(T)$ and $\exp\lp \imag s (E_\mathrm{trial} - H) \rp$, evolving states forward in time.
This puts a bias on the randomized circuits, as a specific Pauli exponential that is drawn from the Hamiltonian will always be evolved with the same sign in the time evolution.
This is not an issue in a noiseless circuit, but may allow systematic errors to appear when noise is present, e.g. due to calibration errors.
To mitigate this bias, we alternate the direction of time when generating circuits, and adapt the sign of $\rho$ accordingly.

\paragraph{Parity post-selection.}
The conservation of parity throughout each circuit can further be used to detect and discard erroneous measurement outcomes.
After measuring the physical qubits (those representing molecular orbitals), any measured state that is in the wrong parity sector indicates an error, and the corresponding shot is discarded.
We refer to this as \textit{parity post-selection}. \par%
\paragraph{Hartree-Fock projection.}
On average, the circuits $U_1^\prime U_2 U_1$ evolve $\initial$ to the ground state of $H$ and back to $\initial$.
This means that even though there will be contributions by states different from $\initial$ in a noiseless setting, on average those contributions will cancel to 0.
Thus, upon measurement, outcomes corresponding to states different from $\initial$ on the non-ancilla qubits can be set to zero in the ancilla's expectation value\ \cite{Granet2024}.
Since $\initial$ is the \ac{HF} state in this work, we call this method \textit{\ac{HF}-projection}. We note that, contrary to violation of parity, we \emph{cannot} post-select away the shots where we do not come back to $\initial$, because even in the noiseless case some shots will differ from $\initial$. However, since we know that they have to average to $0$, we are allowed to set the expectation value to $0$ for these shots.

\section{Results\label{sec:results}}
\subsection{Molecule}
The system we study in this work is the \ce{H3+} molecule.
Its three hydrogen atoms are positioned at the vertices of an equilateral triangle with bond length $1.5\,$\AA, compare \autoref{fig:workflow}.
The electronic structure Hamiltonian is calculated in the STO-3G basis using \texttt{InQuanto}~\cite{inquanto} and its extension of \texttt{PySCF}~\cite{sun2018pyscf}, exploiting $C_{2v}$ point group symmetries.\par%
The molecular Hamiltonian is mapped to a six-qubit Pauli Hamiltonian via the \ac{JW} mapping.
Its interaction norm $\mu_I \approx 2.18$ is reduced to $\mu_I \approx 0.93$ by optimizing \autoref{eq:param_hamiltonian}.
From this point onward, only the optimized Hamiltonian is considered.
Excluding the identity term, the Hamiltonian contains 41 non-trivial Pauli operators with an average Pauli weight of $3.02 \pm 1.18$.
The explicit form of the Hamiltonian is given in \autoref{eq:hamiltonian_numeric} in the Appendix.\par%
The ground state of the initial Hamiltonian $\Hbg$ is the \ac{HF} state, which for this system is $\initial = \ket{110000}$, corresponding to a total charge of $+1$ and two electrons.
The orbitals here are ordered ascendingly from left to right.

\subsection{Algorithm settings}
The adiabatic state preparation $\mathcal{A}(T)$ is performed with a linear sweep function and total evolution time $T = 8\,\mathrm{Ha}^{-1}$ ($\hbar=1$), yielding a state within chemical accuracy of the exact ground-state energy.
The total cost of the experiment depends primarily on the number of circuits, the number of shots, and the number of \acp{TQG} per circuit.
To reduce cost, each two circuits are executed in parallel, which is possible since each requires one ancilla and six physical qubits, and Quantinuum's H1-1 machine is a 20-qubit device.
We will refer to two individual circuits executed in parallel as a stitched circuit.
The trial energies $E_\mathrm{trial}^\pm$ are chosen using the fact that the HF energy $E_\mathrm{HF}$ is an upper bound to the exact \ac{GSE},
\begin{align}
    \lp E_\mathrm{trial}^+, E_\mathrm{trial}^- \rp = \lp E_\mathrm{HF}, E_\mathrm{HF} - 2\epsilon \rp, \quad \epsilon = 0.04\,\mathrm{Ha}.
\end{align}
Here we have chosen $\epsilon=0.04\,\mathrm{Ha}$ since \ac{HF} usually approximates the \ac{GSE} up to a few tens of mHa, and $E_\mathrm{HF} - 0.08\,$Ha is thus likely a lower bound to $E_\mathrm{GS}$~\cite{jones2022chemistry,liu2024perturbative}.
We note that running the algorithm will show whether $E_\mathrm{trial}^-$ is below $E_\mathrm{GS}$, and the procedure can be repeated with a lower trial energy if necessary.

With the given parameters, all that is left is to find suitable values of central time $s$ and gate angle $\tau$.
In principle, any pair of values $(s,\tau)$ would suffice as long as the average number of \acp{TQG} is feasible on hardware.
For Quantinuum's H1-1 trapped-ion quantum computer used in this work, we choose to restrict the average number of \acp{TQG} per individual circuit to be below 1100.
Apart from upper-bounding the average number of \acp{TQG}, we can further motivate a specific choice of $(s,\tau)$ by calculating the variance on the final energy estimate of a set of circuits generated with this setting, and choosing a pair of values such that this variance is minimized.
The result of this analysis is that we choose $(s,\tau) = (10\,\mathrm{Ha}^{-1}, 0.1)$, and the data that supports this choice is shown in \autoref{fig:choice_s_tau}.\par
\begin{figure}[ht!]
    \centering
    \includegraphics[width=\linewidth]{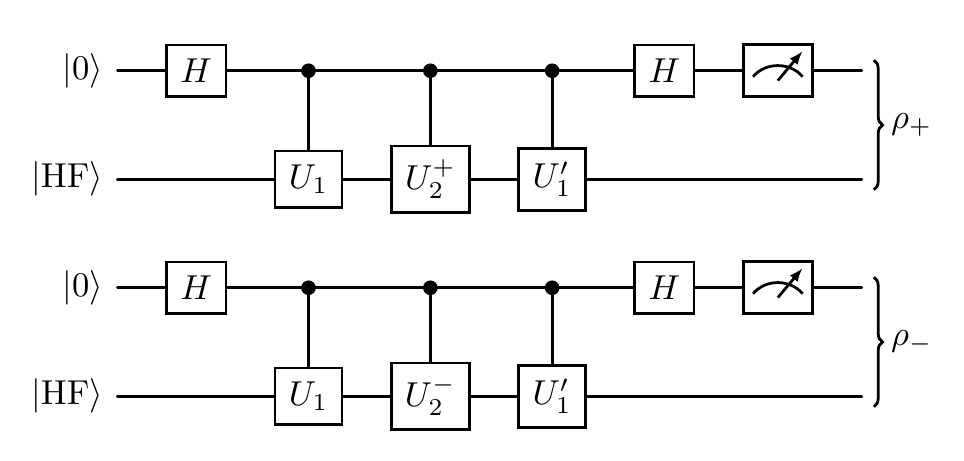}
    \caption{\textbf{Circuits executed on hardware.}
    The stitched circuits run on hardware are comprised of two independent parts measuring $\rho_\pm$.
    We generate 346 distinct circuits of this kind, and measure each one 5 times.
    To reduce variance, $U_1$ and $U_1^\prime$ are identical for $\rho_+$ and $\rho_-$ in every distinct stitched circuit.
    $U_2^+$ and $U_2^-$ only differ by a single-qubit $Z$ rotation on the ancilla that implements the phase shift introduced by $E_\mathrm{trial}^\pm$.
    Measurements on the physical qubits are used for post-selection and not shown in this figure.
    }
    \label{fig:circuit_stacked}
\end{figure}

%
With the setup complete, we generate 346 distinct stitched circuits, shown in \autoref{fig:circuit_stacked}.
Each circuit is measured 5 times, where the circuits for estimating $\rho_+$ and $\rho_-$ are executed in parallel.
Since $\rho_+$ and $\rho_-$ are combined to estimate the \ac{GSE}, one energy estimate will refer to one stitched circuit being measured 5 times.
All 346 energy estimates will be referred to as the hardware experiment.
To further reduce variance, the circuits for $\rho_+$ and $\rho_-$ are identical except for the single-qubit ancilla rotation implementing the phase shift for the respective trial energy.
A simulation of the hardware experiment using noiseless circuits and 5 measurements per circuit yields a final energy estimate of
\begin{align}
    E_\mathrm{noiseless} - E_\mathrm{GS} = -3.3\pm 1.4\,\mathrm{mHa},\label{eq:noiseless_hardware_experiment}
\end{align}
with HF-projection applied to the results.
This confirms that the prepared circuits indeed prepare the ground state and read out its energy with high precision.
In fact, in the limit of infinitely many measurements, the 346 distinct circuits would yield an energy estimate of $E_\mathrm{noiseless} - E_\mathrm{GS} = -0.8\,$mHa, see \autoref{sec:error_analysis}.

We can further gauge the quality of the randomized circuits by evaluating the fidelity of the states that are prepared on average.
Let $\ket{\psi_{U_1}}$ and $\ket{\psi_{U_1^\prime}}$ denote the states that are on average implemented by the respective unitaries,
\begin{align}
    \ket{\psi_{U_1}} &\coloneq \mathcal{N}\sum_{i=1}^{346} U_{1,i} \initial\\
    \ket{\psi_{U_1^\prime}} &\coloneq \mathcal{N^\prime}\sum_{i=1}^{346} \lp U_{1,i}^\prime\rp^\dagger \initial,
\end{align}
where $U_{1,i}^{(\prime)}$ is one randomized implementation of $\mathcal{A}(T)^{(\dagger)}$, and $\mathcal{N}^{(\prime)}$ normalizes the state.
Computing the fidelity to the exact ground state $\ket{\mathrm{GS}}$, $\mathcal{F}(\ket{\phi}) \coloneq  |\langle{\mathrm{GS}}|\phi\rangle|^2$ for these averaged states then yields
\begin{align}
    \mathcal{F}(\ket{\psi_{U_1}}) &\approx 0.999\\
    \mathcal{F}(\ket{\psi_{U_1^\prime}}) &\approx 0.999.
    \intertext{A comparison to the state $\ket{\psi(T)}=\mathcal{A}(T)\initial$ yields $\mathcal{F}(\ket{\psi(T)}) \approx 0.998$, showing that the randomized states' fidelities are close to their exact values.
    The prepared states thus average well to the exact ground state of $H$, and importantly have a higher fidelity than the \ac{HF} state,}
    \mathcal{F}(\ket{\mathrm{HF}}) &\approx 0.947.
\end{align}

In order to assess the resource requirements of the randomized time evolution we use here, we compare it to standard Trotterization and provide estimates of how much deeper circuits would need to be in order to achieve similar algorithmic precision, see \autoref{app:trotter}.
\subsection{Hardware results}
The algorithm is executed on Quantinuum's 20-qubit H1-1 quantum computer~\cite{quantinuumh1}, which features all-to-all connectivity and an average two-qubit gate infidelity of $2.05(7) \times 10^{-3}$~\cite{quantinuumperformance}.
Statistical uncertainties in $\rho_\pm$ are obtained directly from the measurement variances, and these uncertainties are used to numerically determine the uncertainties on the energy estimates.\par%
The main result is shown in \autoref{fig:energy_convergence_variance}.
The accumulated energy estimate $E_i$ denotes the energy estimate using all measurements up to the $i$-th, and evaluates to the final estimate
\begin{align}
    E_\mathrm{final} - E_\mathrm{GS} = -25.5\pm 5.3\,\mathrm{mHa}.\label{eq:E_final}
\end{align}
This value is obtained after applying parity post-selection and HF projection.
For comparison, the classical HF method yields $E_\mathrm{HF} - E_\mathrm{GS} = 52.8\, \mathrm{mHa}$, demonstrating that the quantum algorithm improves the absolute deviation from the true \ac{GSE} by more than $5$ standard deviations. The \ac{HF} energy is shown for reference only and is not included in $E_i$.
The small energy error of 5.3\,mHa indicates that the deviation from the noiseless value is mostly because of hardware noise rather than shot noise, cf. \autoref{eq:noiseless_hardware_experiment}.
We will discuss the effects of hardware noise later in \autoref{sec:error_analysis}.\par
\begin{figure*}[ht!]
    \centering
    \includegraphics[width=\linewidth]{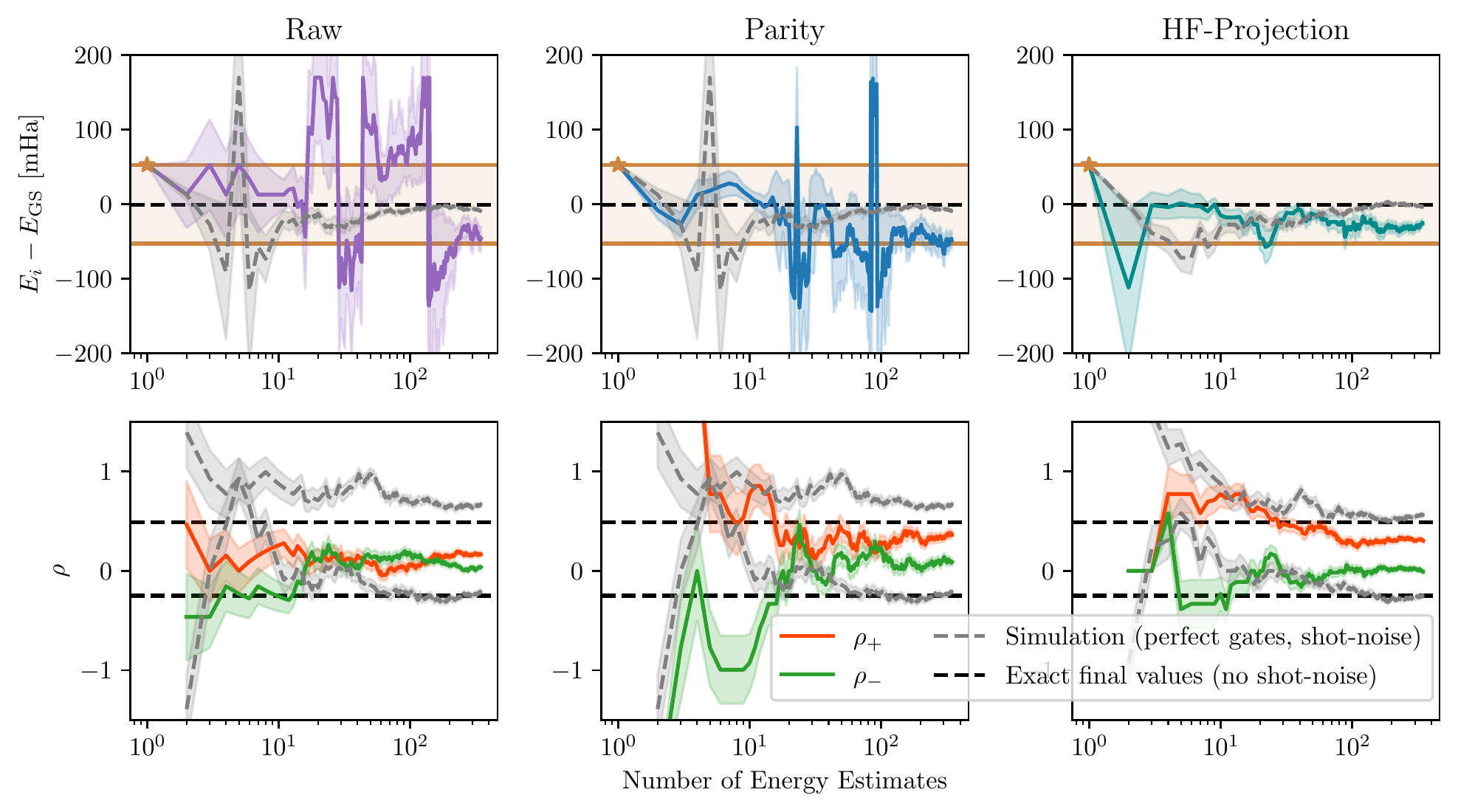}
    \caption{\label{fig:ps_modes_energy_rho_convergence}\textbf{Effect of post-selection on hardware results.}
    \textit{Top panel:}
    Energy estimate for raw data (purple), parity post-selection (blue), and parity \& HF projection (teal).
    Shaded regions on curves indicate statistical uncertainty, and the orange line shows $E_{\mathrm{HF}}$ (for reference only).
    \textit{Bottom panel:}
    Accumulated $\rho_+$ (orange) and $\rho_-$ (green) values.
    Gray curves in both panels show simulations with perfect gate execution, but the same number of shots as in the hardware experiment.
    Horizontal black lines in both panels indicate the exact final values one would obtain with the given circuits, executed perfectly and without shot noise.
    One energy estimate corresponds to 5 single-shot measurements of a circuit depicted in \autoref{fig:circuit_stacked}, some of which are removed through post-selection.
    }
\end{figure*}

%
The values in \autoref{fig:energy_convergence_variance} and \autoref{eq:E_final} are obtained after applying parity post-selection and \ac{HF} projection.
To isolate the effect of these post-processing steps, the top panel of \autoref{fig:ps_modes_energy_rho_convergence} illustrates the results for three modes: \textit{Raw} (no post-selection), \textit{Parity} (parity post-selection), and \textit{HF-Projection} (parity post-selection followed by HF projection).
The corresponding values of $\rho_\pm$, which are the ancilla expectation values used to estimate $E_i$, are shown in the bottom panel, and both panels include a noiseless simulation of the same experiment.
The noiseless simulation is not affected by parity post-selection because the correct parity sector only changes when errors occur; yet the \ac{HF} projection still has an effect due to the finite number of measurements.\par
\begin{figure}[ht!]
    \centering
    \includegraphics[width=\linewidth]{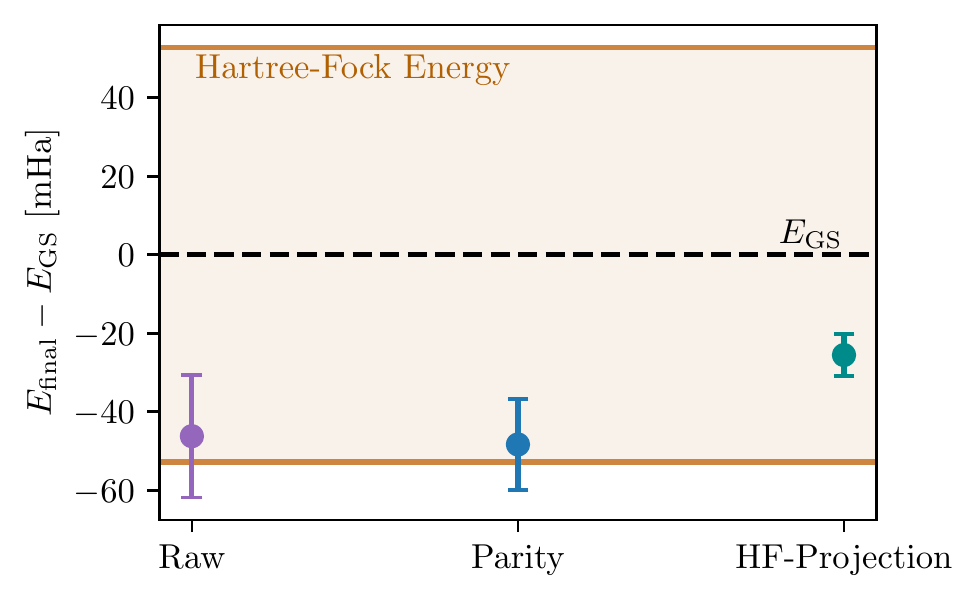}
    \caption{\label{fig:final_estimates_psmodes}\textbf{Final energy estimates of raw and post-selected data.}
    The data points show the energy estimate for different post-selection methods.
    HF-projection, which includes parity post-selection, clearly enhances the quality of the estimate, making the results statistically incompatible to the HF energy.
    }
\end{figure}

%
Parity post-selection discards $\sim 62\,\%$ of measurements, removing data known to contain errors.
This step helps maintain $\rho_+ > \rho_-$, visibly preventing sign flips in $E_i$.
\ac{HF}-projection does not discard runs but sets ancilla values that should average to zero directly to zero, improving stability: $\rho_+ > \rho_-$ is always satisfied, and no abrupt jumps occur.
\autoref{fig:final_estimates_psmodes} shows the final estimates for the three post-processing modes.
Only the \ac{HF}-projected result is statistically distinguishable from the classical \ac{HF} approximation, underlining the effectiveness, but also necessity of classical post-selection techniques.\par%
Investigating the data further, we examine the values of $\rho_+$ and $\rho_-$ after \ac{HF}-projection in \autoref{tab:rho}.
\begin{table}[ht!]
    \centering
    \caption{\label{tab:rho}Ancilla expectation values $\rho_\pm$ after HF-projection.}
    \begin{tabular}{l | c | c}
    \toprule
            & $\rho_+$ & $\rho_-$ \\\midrule
        Hardware & $0.302\pm 0.022$ & $-0.007\pm 0.023$ \\
        Noiseless & $0.441 \pm 0.019$ & $-0.220 \pm 0.019$ \\\bottomrule
    \end{tabular}
\end{table}
Comparing hardware to noiseless results, the damping factors $f_\pm = \rho_\pm/\rho_\mathrm{\pm,noiseless}$ that quantify how much the hardware signal is damped compared to the noiseless setting are $f_+ \approx 0.68$ and $f_- \approx 0.03$.
The large asymmetry of $f_+$ and $f_-$ skews the ratio $\frac{\rho_+ + \rho_-}{\rho_- - \rho_+}$ in \autoref{eq:GSE_estimate}, leading to a biased GSE estimate.
In \autoref{subsec:energy_estimation} we argued that depolarizing noise should yield similar damping factors $f_\pm$, making the estimate robust against this type of noise.
The observed asymmetry, however, suggests other error sources, which we will analyze in \autoref{sec:error_analysis}.

\section{Error analysis\label{sec:error_analysis}}
To identify the dominant error sources responsible for the deviations between the ideal simulation and the experimental results, we perform a series of noisy simulations and compare them to the hardware data.

\subsection{Incoherent and Coherent Noise}
\begin{figure}[ht!]
    \centering
    \includegraphics[width=\linewidth]{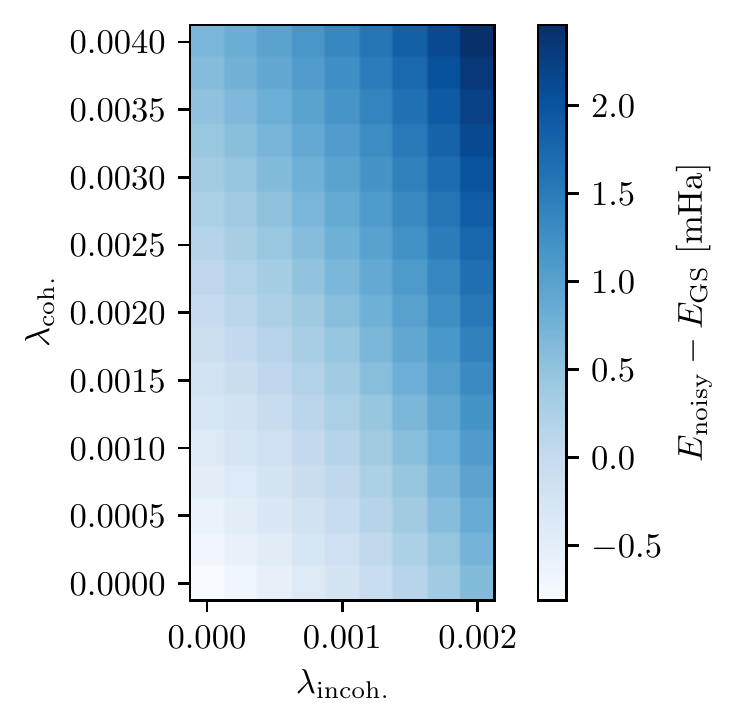}
    \caption{\label{fig:density_matrix_energy}\textbf{Noisy density matrix simulation.} Each data point shows the energy estimate of a noisy density matrix simulation containing the circuits that were submitted to hardware.
    The values are obtained after HF-projection.
    The effect of incoherent and coherent errors on the energy estimate remains small, even though the simulated error rates go beyond their reported hardware values.
    }
\end{figure}

We first consider incoherent and coherent noise sources.
The simulations are performed by approximating the two-qubit gate error with an incoherent two-qubit error channel, and using a single-qubit coherent error channel to mimic memory error per depth-1 circuit time~\cite{quantinuumperformance}.
The two-qubit incoherent error channel is defined as
\begin{align}
    \mathcal{E}_\mathrm{incoh.} [\rho] = (1-\lambda_\mathrm{incoh.})\rho + \lambda_\mathrm{incoh.} \tr (\rho) \frac{I}{4},
\end{align}
for $\rho$ a density matrix and where the trace and the identity operator act on the two-qubit sub-space, while the single-qubit coherent error channel is defined as
\begin{align}
    \mathcal{E}_\mathrm{coh.} [\rho] = (1-\lambda_\mathrm{coh.}) + \lambda_\mathrm{coh.} R_Z \rho  R_Z^\dagger,
\end{align}
with $R_Z = \mathrm{exp}\lp -\imag \frac{\pi}{8} Z \rp$ a single-qubit $Z$ rotation. $\lambda_\mathrm{incoh.},\ \lambda_\mathrm{coh.}$ are the incoherent and coherent error rates.
The resulting error channel is then defined as
\begin{align}
    \mathcal{E} [\rho] = \lp \mathcal{E}_\mathrm{incoh.} \circ \mathcal{E}_\mathrm{coh.} \rp [\rho],
\end{align}
and the noisy energy expectation value is estimated via
\begin{align}
    E_\mathrm{noisy} = \tr\lp\mathcal{E}[\rho] H\rp.
\end{align}
\autoref{fig:density_matrix_energy} shows results from noisy density matrix simulations, where we vary the depolarizing and coherent error rates beyond their nominally reported hardware values of $\lambda_\mathrm{coh.}^\mathrm{H1-1}\approx 2.2\times 10^{-4}$ and $\lambda_\mathrm{incoh.}^\mathrm{H1-1}\approx 9.7\times 10^{-4}$~\cite{quantinuumperformance}.
The data shows the deviation of $E_\mathrm{noisy}$ from the exact value $E_\mathrm{GS}$ with HF-projection applied to the numerical samples.
The data point in the bottom left corner shows the energy estimate in the noiseless case, i.e. without circuit or shot noise, of about $-0.8$\,mHa.
The reason for this value not being exactly zero is that there is a finite number of randomized circuits, although each of the circuits is estimated using an effectively infinite number of shots in the density matrix simulation.

Even under the most pessimistic conditions considered here (i.e. $\lambda_\mathrm{incoh.} = 2\times 10^{-3}$, $\lambda_\mathrm{coh.} = 4\times 10^{-3}$) the resulting deviation in the estimated energy remains small, on the order of $2$\,mHa.
This corresponds roughly to the threshold of chemical accuracy and is an order of magnitude smaller than the deviation observed in the hardware experiments.
These findings indicate that the algorithm is resilient against incoherent and coherent noise, and that these sources of noise are insufficient to explain the deviation in energy of the hardware run.

\subsection{Leakage Errors}
We next turn to simulating leakage errors, which we implement by assigning leakage events to two-qubit operations with probability $\lambda_\mathrm{leakage}$.
We approximate leakage errors with the following model:
If a leakage event occurs, all subsequent operations that include actions on the leaked qubit are removed from the circuit.
Since leaked qubits are detected as $\ket{1}$, we reset the qubit to $\ket{1}$ after it has leaked.

\begin{figure}[ht!]
    \centering
    \includegraphics[width=\linewidth]{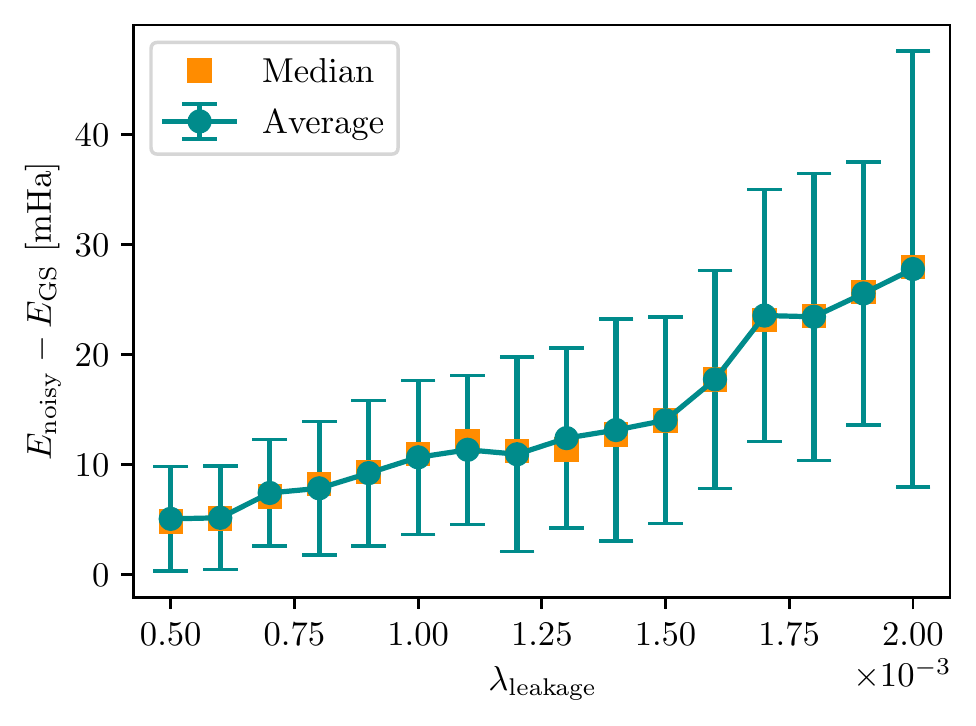}
    \caption{\label{fig:leakage_error_scaling}\textbf{Leakage error scaling.} Each data point shows the average and median energy estimate for 150 independent noisy simulations of the experiment run on hardware, with HF-projection applied.
    Error bars indicate one standard deviation.
    The systematic bias introduced by leakage has a significant effect on the energy estimate.}
\end{figure}

The effects of applying this noise model are illustrated in \autoref{fig:leakage_error_scaling}.
Each data point is obtained by independently simulating the hardware experiment 150 times and applying HF-projection to the measurement outcomes.
As the leakage probability increases, a systematic bias on the energy estimate emerges.
The main point is that the offset here is much larger than the one introduced by coherent and incoherent errors combined.
In fact, the energy estimate on hardware is $E_\mathrm{final} - E_\mathrm{GS} \approx - 25.5$\,mHa, meaning the offset introduced by leakage errors is of the same order of magnitude.
This sensitivity to leakage errors, together with the low sensitivity to coherent and uncoherent noise, shows that leakage is likely to be responsible for the larger deviations observed in the hardware data.

The leakage error rates considered in our simulations exceed those reported for Quantinuum's H1-1 device, which are on the order of $2.2\times 10^{-4}$~\cite{quantinuumperformance, chen2025randomized}.
However, this choice is justified by several considerations.
First, leakage processes often depend strongly on circuit depth, and the circuits used here are considerably deeper than those used in calibration experiments~\cite{chen2025randomized}.
We provide some statistics on the circuits and their depths in \autoref{app:circuit_statistics} of the appendix.
Second, leakage manifests itself as a bias toward measuring more qubits in the $\ket{1}$ state, since leaked qubits are typically read out as $\ket{1}$.
This effect is clearly visible in the hardware measurements. Third, leakage error benchmarking is typically done for dense circuits in which shuttling and gating affects all qubits approximately evenly. The Hadamard-test structure of our algorithm imposes a one-to-many structure in which one qubit (the ancilla) is shuttled and gated disproportionately. Leakage of the ancilla out of the computational subspace in such a circuit is clearly a catastrophic event and we will now go on to show that a large part of the error is caused by such events.

\subsection{Signature of Leakage in Measurement Statistics}
\begin{figure*}[ht!]
    \centering
    \begin{subfigure}[t]{0.49\textwidth}
        \centering
        \caption{\label{fig:leakage_bits_phys}\textit{Physical qubits.}}
        \includegraphics[width=\linewidth]{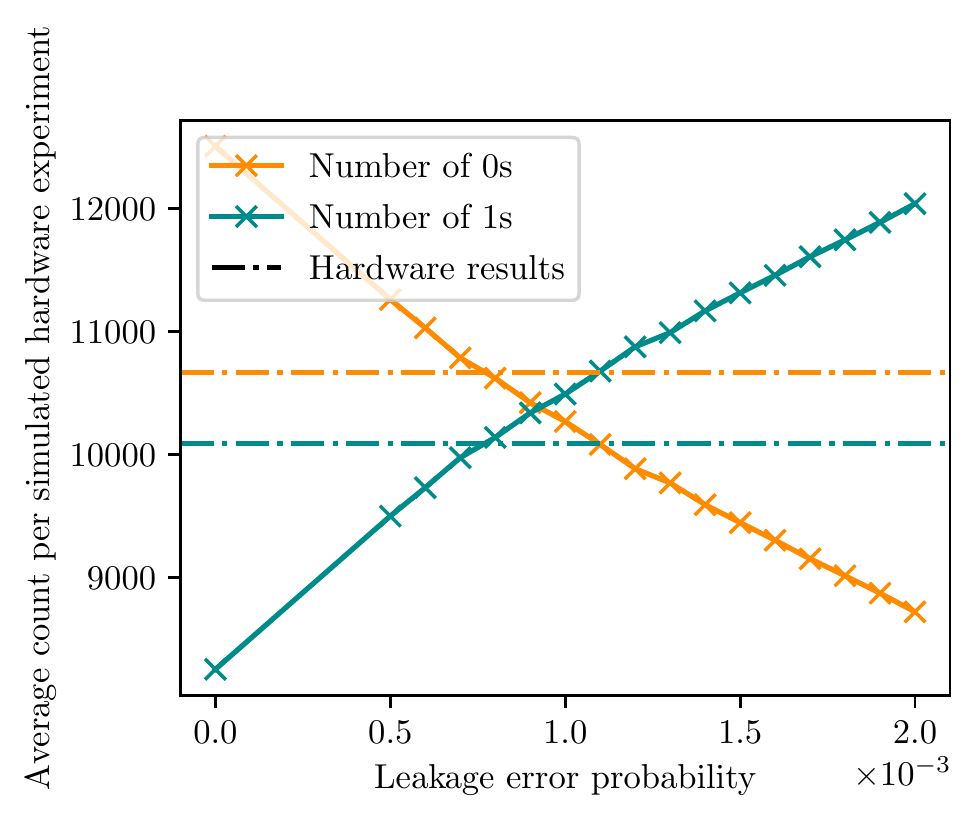}
    \end{subfigure}%
    ~
    \begin{subfigure}[t]{0.49\textwidth}
        \centering
        \caption{\label{fig:leakage_bits_anc}\textit{Ancilla qubit.}}
        \includegraphics[width=\linewidth]{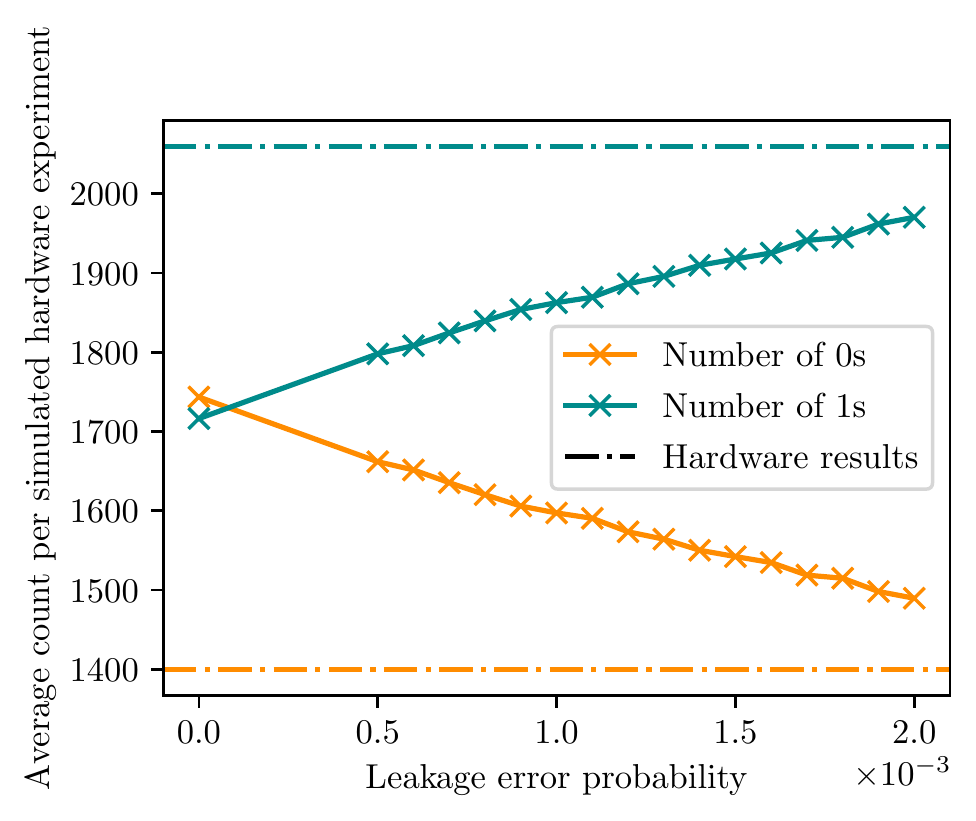}
    \end{subfigure}
    \caption{\label{fig:choice_s_tau}\textbf{Signature of leakage in measurement statistics.}
    The data shown in these graphs indicates the presence of large leakage probabilities.
    Each data point shows the mean number of 0s and 1s measured in noisy simulations, averaged over 150 independent repetitions of the entire experiment.
    No post-selection is applied to directly access the effect of leakage, as leaked qubits are typically measured in the $\ket{1}$ state.
    The horizontal lines indicate the number of times 0s and 1s were measured in the hardware run.
    Figure (a) shows the average number of 0s and 1s on the physical qubits, indicating that leakage error rates could be on the order of $0.6\times 10^{-3}$ per two-qubit gate for our circuits.
    Figure (b) shows the same for the ancilla qubit.
    The imbalance of 0s and 1s compared to the noiseless case is even more pronounced, indicating that leakage error rates could be even larger.
    }
\end{figure*}

To quantify this bias, \autoref{fig:leakage_bits_phys} shows the average number of measured $\ket{0}$ and $\ket{1}$ outcomes for the physical qubits as a function of leakage probability.

The data points are obtained from exactly the same data as in \autoref{fig:leakage_error_scaling}, complemented by another data point in a noiseless setting.
This time, however, no post-selection is applied in order to access the immediate effect of leakage on the measurement outcomes.
The hardware data exhibits a significantly larger number of $\ket{1}$ outcomes than expected from noiseless simulations, consistent with the presence of leakage.
The simulated leakage levels required to reproduce this imbalance are comparable to the lower end of values in~\autoref{fig:leakage_error_scaling}.

For the ancilla qubit, see \autoref{fig:leakage_bits_anc}, the discrepancy is even more pronounced.
The hardware results exhibit a strong bias toward $\ket{1}$ outcomes that can only be explained by assuming even higher leakage rates than those suggested for the non-ancilla qubits.
Based on this analysis, we conclude that leakage occurring at higher rates is very likely the dominant contributor to the experimental deviations.

\section{Discussion\label{sec:discussion}}
In this work we demonstrate practical ground state energy estimation on an ion-trap quantum computer, improving on the classical \ac{HF} estimate despite noise.
We perform classical pre-processing techniques to shorten circuit depths, and apply two types of post-selection to mitigate noise.
The hardware results are complemented by a thorough error analysis, obtained through classical simulations.
We show that while the algorithm is resilient to coherent and incoherent noise, there remains a strong sensitivity to leakage errors.
The noisy simulations suggest that the energy estimate on hardware would have yielded much better results if no leakage had occurred, achieving a precision on ground state energy that would be just twice chemical accuracy, even with shot noise included.
This analysis sparks several potential avenues.

First, further improvements would help to reduce the interaction norm $\mu_I$ and circuit depths. For example, double factorization methods could be employed to find decompositions of the Hamiltonian that have a lower norm \cite{cohn2021quantum,oumarou2024accelerating}. There are also representations of chemistry Hamiltonians in an enlarged basis set that could prove more resource-sparing for the norm \cite{luo2025efficient}. There are finally other fermionic encodings such as the ternary tree encoding that have better asymptotic scalings than Jordan-Wigner and that would reduce circuit depth through the implementation of rotation of shorter Pauli strings \cite{jiang2020optimal}.

Second, the strong dependence of the efficiency of the algorithm on leakage of the ancilla qubit motivates efforts towards mitigating its influence.
One could e.g. encode the ancilla into a leakage error detection code that is periodically checked for leakage, and suppress leakage errors by converting them to Pauli errors~\cite{fowler2013coping}.
A leakage detection gadget at the end of the circuit further allows to post-select shots where the ancilla qubit has leaked. Given the potentially significant improvement that the mitigation of leakage error would give on the ground state energy accuracy, as argued in this work, this is a promising direction.

\section*{Acknowledgements}
We thank Duncan Gowland and David Zsolt Manrique for their valuable feedback on this manuscript.

This work received support from the German Federal Ministry of Education and Research via the funding program quantum technologies - from basic research to the market under contract number 13N16067 “EQUAHUMO”.
It is also part of the Munich Quantum Valley, which is supported by the Bavarian state government with funds from the Hightech Agenda Bayern Plus.
E.G. acknowledges support by the Bavarian Ministry of Economic Affairs, Regional Development and Energy (StMWi) under project Bench-QC
(DIK0425/01). 

\section*{Data Availability}
Data and code were obtained using \texttt{InQuanto} code, which is proprietary resource.

\begin{acronym}

\acro{ASP}{adiabatic state preparation}

\acro{CASSCF}{complete active space self-consistent field}
\acro{GSE}{ground-state energy}
\acroplural{GSE}[GSEs]{ground-state energies}

\acro{HF}{Hartree-Fock}

\acro{iQPE}{iterative quantum phase estimation}

\acro{JW}{Jordan-Wigner}

\acro{NISQ}{noisy intermediate-scale quantum}

\acro{QPE}{quantum phase estimation}

\acro{TQG}{two-qubit gate}
\acroplural{TQG}[TQGs]{two-qubit gates}

\end{acronym}

\bibliography{literature.bib}

\appendix
\null
\section{Time Evolution Through Random Independent Sampling\label{app:tetris}}
We present an outline of the randomized time evolution used in this work, and refer to Refs.\ \cite{Granet2023, Granet2024} for further details.
The algorithm starts by defining the monotonically increasing function $z(u)$,
\begin{align}
    z(u) = \int_0^u w(u^\prime)\mathrm{d}u^\prime,
\end{align}
along with its reciprocal function $z^{-1}[z(u)]=u$ for all $u$.
Here we will restrict $0\leq u \leq 1$.
Let us further define $\zeta = z(1)$ which is of order $\mathcal{O}(1)$.
Choosing a gate angle $0<\tau<\pi/2$ and denoting $\ket{\psi}$ as the state of the system, the implementation of $\mathcal{A}(T)$ proceeds as follows:
\begin{enumerate}
    \item Initialize $\ket{\psi}$ to $\initial$.
    \item For $n=1,\dots,N_I$, draw an integer $m_n\geq 0$ from a Poisson distribution with parameter $\frac{|a_n|\zeta T}{\sin \tau}$.
    \item For $n=1,\dots,N_I$, draw $m_n\geq 0$ real numbers $\Tilde{t}_{i,n}$ (with $i=1,\dots,m_n$), independently and uniformly at random between 0 and $\zeta$. For every $i$, $n$, set $t_{i,n} = Tz^{-1}(\Tilde{t}_{i,n})$.
    \item Find the sequence $(i_1,n_1),\dots,(i_M,n_M)$ such that
    \begin{align}
        o < t_{i_1,n_1} < \cdots < t_{i_M,n_M} < T,
    \end{align}
    where $M=\sum_{n=1}^{N_I}m_n$.
    \item For $m=1,\dots,M$, apply the operator
    \begin{align}
        \exp\lb \imag \tau P_{n_m}\rb \exp \lb \imag \lp t_{i_m,n_m} - t_{i_{m-1},n_{m-1}}\rp H_B\rb
        \label{eq:operator_instance}
    \end{align}
    on $\ket{\psi}$, with initial condition $t_{i_0,n_0}\equiv 0$.
    \item Apply $\exp\lb \imag \lp T - t_{i_m,n_m}\rp H_B \rb$ on $\ket{\psi}$.
\end{enumerate}
Denoting by $U$ the random unitary generated by the algorithm, its statistical average $\mathbb{E}$ is
\begin{align}
    \mathbb{E}[U] = \mathrm{e}^{-\tan(\tau/2)\zeta T \mu_I} \mathcal{A}(T).
\end{align}

\section{Choice of central time and gate angle}
\begin{figure*}[ht!]
    \centering
    \begin{subfigure}[t]{0.49\textwidth}
        \centering
        \caption{\label{fig:choice_s_tau_variance}}
        \includegraphics[width=\linewidth]{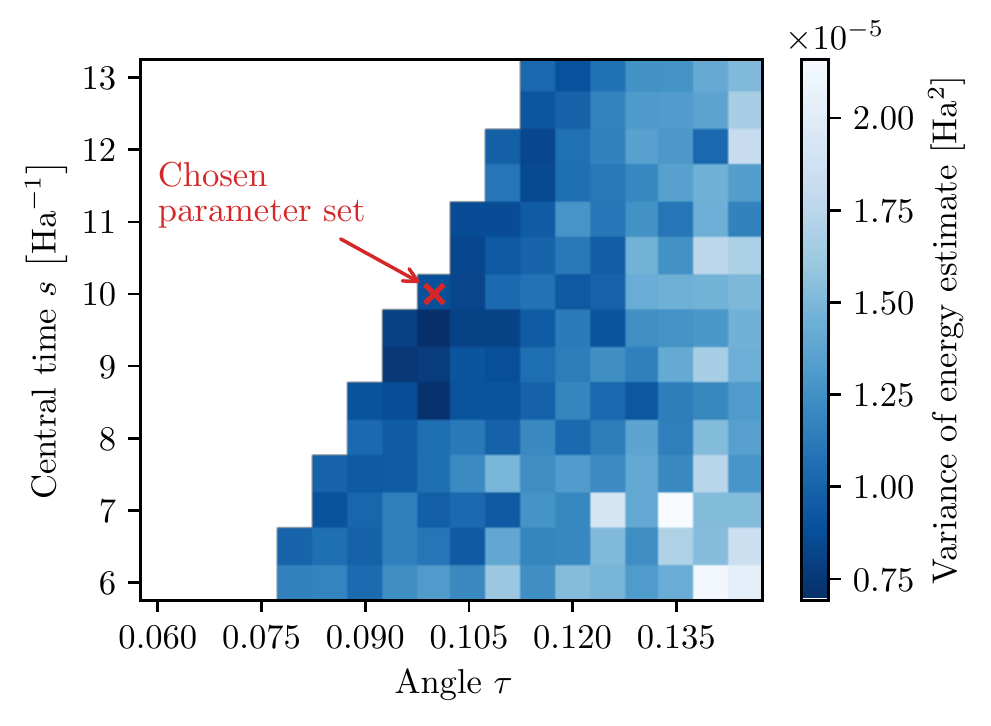}
    \end{subfigure}
    ~
    \begin{subfigure}[t]{0.49\textwidth}
        \centering
        \caption{\label{fig:choice_s_tau_tqg}}
        \includegraphics[width=\linewidth]{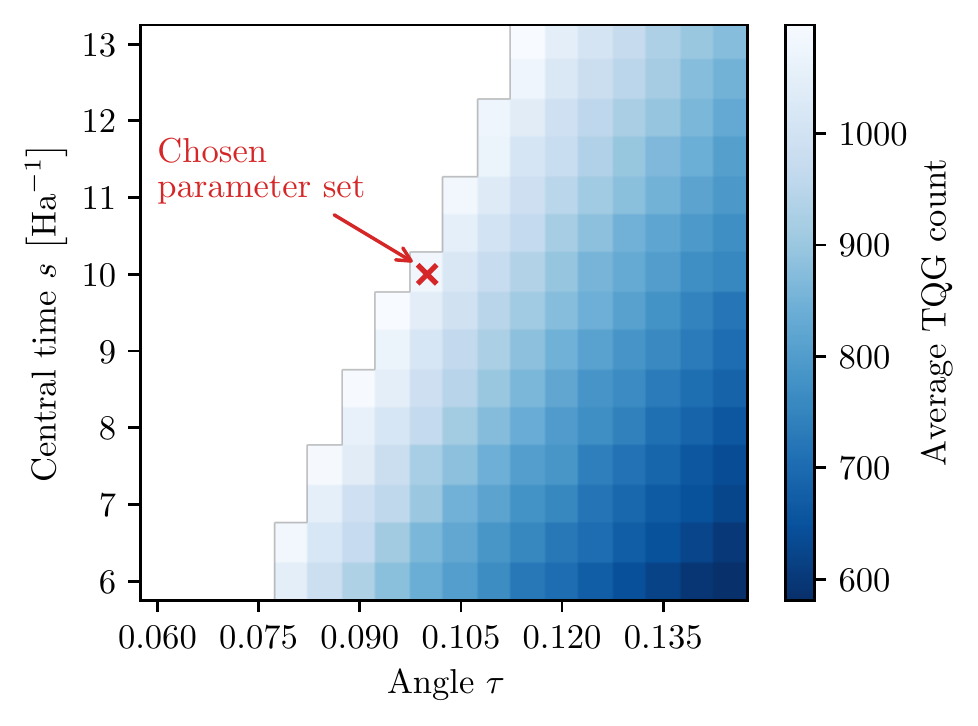}
    \end{subfigure}%
    \caption{\textbf{Average \ac{TQG} count and energy variance.} The data shown in these graphs motivates the coice of central time $s$ and gate angle $\tau$ in this work.
    For each pair $(s,\tau)$ and assuming 5 single-shot measurements, circuits were generated and analyzed.
    Both plots show data for which the total number of \acp{TQG} is below 1100.
    Figure (a) shows the variance of the energy estimate, simulated without shot and circuit noise.
    The final choice of $s=10\,\mathrm{Ha}^{-1}$ and $\tau=0.1$ was made such that the variance is roughly minimized.
    Figure (b) shows the average number of \acp{TQG} per circuit.
    }
\end{figure*}

The central time $s$ and gate angle $\tau$ are parameters that can be chosen relatively freely, at the cost of having to run more or less circuits with fewer or more gates.
Since we are able to simulate the circuits used in this work, we can motivate the choice of $s$ and $\tau$ by determining the variance of the resulting energy estimate numerically.
Due to the fact that we have a noisy quantum device and do not want to execute more than 1100 \acp{TQG} per individual circuit, we restrain the average number of \acp{TQG} to be below that value.
We then simulate a hardware experiment for various pairs of values $(s,\tau)$ that fulfill our constraints, and plot the variance in \autoref{fig:choice_s_tau_variance}.
One can see that our choice of $s=10\,\mathrm{Ha}^{-1},\tau=0.1$ is close to the minimum variance, which enhances the quality of the energy estimate.
The data in \autoref{fig:choice_s_tau_tqg} visualizes the \ac{TQG} count constraint.

\section{Circuit statistics\label{app:circuit_statistics}}
\begin{figure}[ht!]
    \centering
    \includegraphics[width=\linewidth]{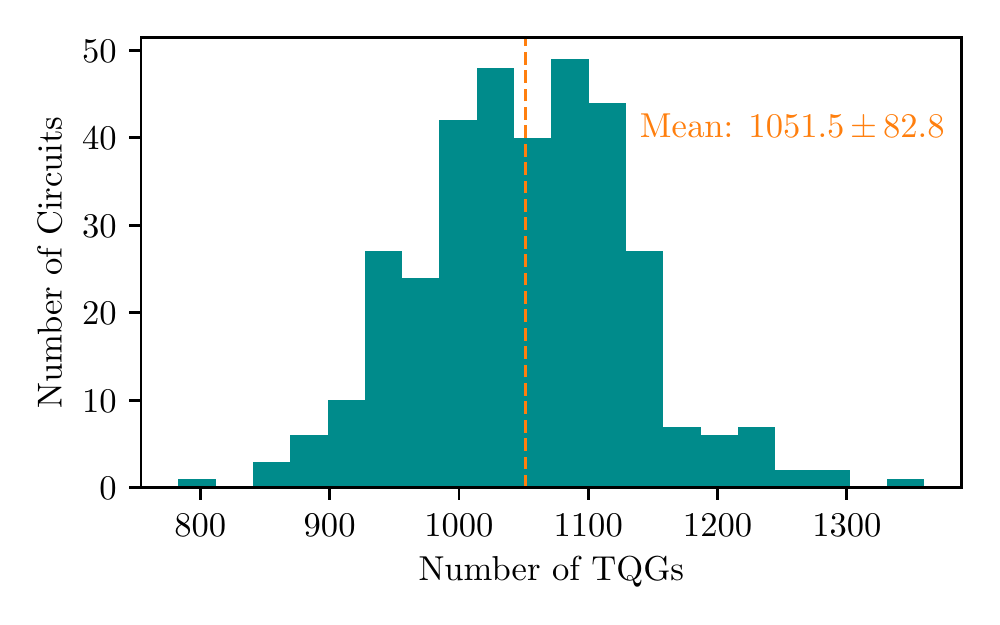}
    \caption{\textbf{Number of \acp{TQG} per circuit.} For all of the $2\times 346$ individual circuits that each estimate $(\rho_+,\rho_-)$, the number of \acp{TQG} is recorded. Since the circuits for estimating $\rho_\pm$ are equivalent except for a single-qubit rotation on the ancilla qubit, only one of the two circuits is considered.}
    \label{fig:tqg_histogram}
\end{figure}

\begin{figure}[ht!]
    \centering
    \includegraphics[width=\linewidth]{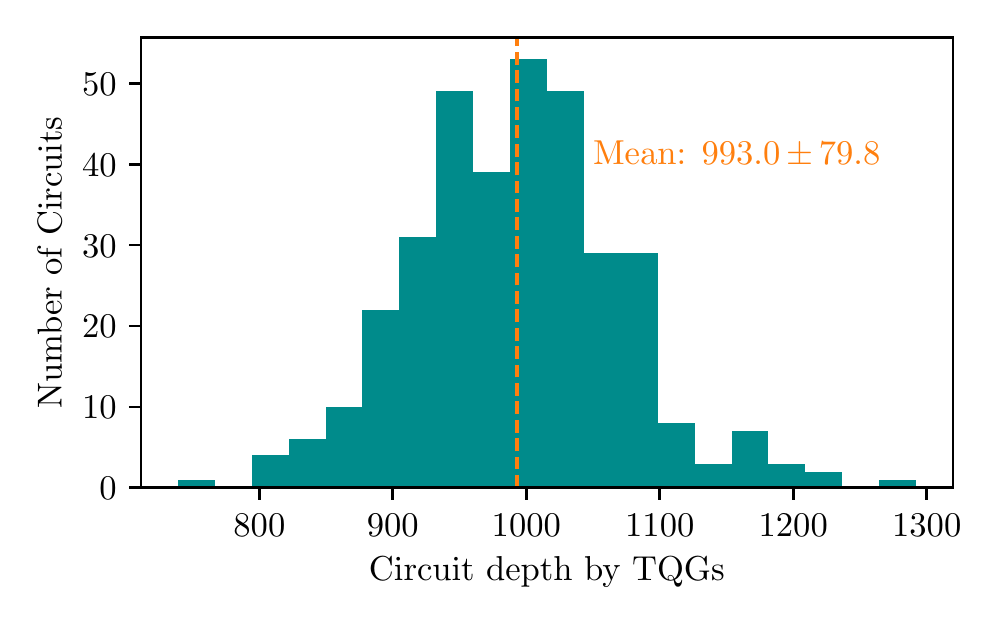}
    \caption{\textbf{Circuit depth by the number of \acp{TQG} per circuit.} For all of the $2\times 346$ individual circuits that estimate $(\rho_+,\rho_-)$, the number of \acp{TQG} is recorded. Since the circuits for estimating $\rho_\pm$ are equivalent except for a single-qubit rotation on the ancilla qubit, only one of the two circuits is considered.}
    \label{fig:depth_histogram}
\end{figure}

We provide some circuit statistics to emphasize the non-triviality of the performed hardware experiment.
\autoref{fig:tqg_histogram} shows the number of \acp{TQG} for each of the $2\times 346$ circuits that estimate $\rho_+$ and $\rho_-$.
Since the circuits for estimating $\rho_\pm$ are equivalent except for a single-qubit rotation on the ancilla qubit, only one of the two circuits is considered.
The mean number of \acp{TQG} per circuit turns out to be $1051.5\pm 82.8$, where the uncertainty corresponds to one standard deviation.
Since we are investigating a quantum chemistry problem in this work, the Jordan-Wigner-mapped Pauli Hamiltonian exhibits non-local terms that span the entire system.
It is therefore unlikely that exponentials of Pauli operators, drawn at random, can be implemented in parallel very often.
This notion is supported by the circuit depth by the number of \acp{TQG}, shown in \autoref{fig:depth_histogram}.
The average circuit depth by the number of \acp{TQG} is $993.0\pm 79.9$, where the uncertainty again corresponds to one standard deviation.
The fact that this quantity is relatively close to the average number of \acp{TQG} in a circuit tells us that operations can indeed only rarely be parallelized due to the Pauli operators' weights.
Additionally, circuit depths on this scale are much deeper than the circuit depths investigated in the analysis of leakage errors on Quantinuum's H1-1 device in Ref.~\cite{chen2025randomized}.
Since leakage is affected by deeper circuits, this analysis strengthens the argument that larger leakage rates can be used to explain the errors occurring in the hardware experiment.

\section{Comparison to Trotterization\label{app:trotter}, iQPE and direct sampling}
\paragraph{Trotterization.}
To assess the performance of the randomized time evolution method, we compare it against conventional Trotterization.
The adiabatic state preparation used in this work, i.e. the operator $\mathcal{A}(T=8\,\mathrm{Ha}^{-1})$, prepares the ground state with an energy expectation deviating from the true \ac{GSE} by $1.49\,$mHa, which is below the threshold for chemical accuracy, defined as $1.52\,$mHa.\par%
Two effects contribute to discretization errors when using Trotterization, namely (i) the decomposition of the adiabatic path into discrete chunks and (ii) the Trotterization of the instantaneous Hamiltonian within each chunk.
To isolate the effect of adiabatic path discretization, we divide the evolution into $k$ time steps of length $T/k$ and compute the product of the resulting operator exponentials,
\begin{align}
    \mathcal{T}\, \mathrm{exp}\lp \imag \int_0^T H\lp t\rp \mathrm{d}t\rp \approx \prod_{a=1}^{k} \mathrm{exp}\lp \imag H\lp a \frac{T}{k}\rp \frac{T}{k}\rp.\label{eq:time_discretized}
\end{align}
Setting $k=4$, this yields an approximation error of $1.58\,$mHa.\par%
In order to implement each term in \autoref{eq:time_discretized}, we consider a first-order Trotter decomposition with $m$ Trotter layers.
The Trotterized evolution is then given by~\cite{suzuki1985decomposition}
\begin{align}
    \exp\left( \imag t \sum_{h\in H(u)} h \right) = \left[\prod_{h\in H(u)} \exp\left( \imag \frac{t}{m} h \right)\right]^m + \mathcal{O}\lp \frac{t^2}{m} \rp,
\end{align}
which increases the circuit depth by a factor of $m$.
Setting $k=4$ and $m=2$, the resulting approximation error is $1.47\,$mHa, which is below the threshold of chemical accuracy.\par%
If each operator with weight greater than two is decomposed using Pauli gadgets, the total number of two-qubit gates required for the Trotterized circuit amounts to 1320.
In contrast, the randomized adiabatic state preparation without discretization error required on average $198.3\pm 33.2$ two-qubit gates.
This shows that the randomized method in this scenario is significantly more efficient in terms of two-qubit gate count.
\paragraph{Direct sampling.}
We now want to estimate the total number of samples needed if we wanted to determine the ground state energy by measuring the individual terms in the Hamiltonian.
To do so, we assume that, for each shot, one of the Paulis in the Hamiltonian is chosen with a probability that is proportional to the absolute value of its coefficient.
We then perform a single-shot measurement in this basis and compute the expectation value from this.
We find that in order to determine the \ac{GSE} with a precision of about $1.6\,$mHa, we would need roughly $9\times 10^6$ single-shot measurements.
This is a very large number of measurements, recalling that the randomized scheme we use in this work needs $2\times 5\times 346=3460$ single-shot measurements to get to similar precision.
\paragraph{\ac{iQPE}.}
For a comparison to \ac{iQPE}, we evaluate the spectrum of a first-order Trotterization of the Hamiltonian,
\begin{align}
    U_\mathrm{iQPE} = \prod_{h\in H(1)}\exp\lp \imag \tau h\rp,
\end{align}
and choose the time step $\tau$ such that $U_\mathrm{iQPE}$ produces the eigenvalue $\exp(\imag \tau E_\mathrm{iQPE})$, where $E_\mathrm{iQPE}$ is within chemical accuracy of the true \ac{GSE}.
For the iterative phase estimation procedure, controlled-$(U_\mathrm{iQPE})^{2^l}$ operations then need to be implemented on the quantum computer to estimate the ground state energy with precision $2^{-(l+1)}/\tau$.\par%
For the $\ce{H_3^+}$ molecule studied in this work, we find that setting $\tau=0.4\,\mathrm{Ha}^{-1}$ yields an eigenvalue $E_\mathrm{iQPE}-E_\mathrm{GS} = 1.4\,$mHa, which satisfies the required chemical accuracy.
A precision of $2^{-(l+1)}/\tau = 1.2\,$mHa is then reached by setting $l=10$.
If we again calculate the number of \acp{TQG} needed to implement these circuits, we find that we would need about 422,000 \acp{TQG} across all circuits, and the maximum number of \acp{TQG} in a single circuit would be approximately 211,000.
This number is of course too large for current noisy devices, showing that the randomized approach we use in this work is the best choice out of the three in this scenario.

\section{Pauli Hamiltonian}
The Hamiltonian after optimizing \autoref{eq:param_hamiltonian} is given in \autoref{eq:hamiltonian_numeric}.
\begin{widetext}
\begin{align}
\label{eq:hamiltonian_numeric}
\begin{aligned}
H=&-2.77 IIIIII\\&
+5.81\times 10^{-1} IIIIIZ
+5.81\times 10^{-1} IIIIZI
+5.81\times 10^{-1} IIIZII
+5.81\times 10^{-1} IIZIII\\&
+7.48\times 10^{-1} IZIIII
+7.48\times 10^{-1} ZIIIII
+2.46\times 10^{-2} IIIIZZ
-2.01\times 10^{-3} IIIZIZ\\&
-1.87\times 10^{-2} IIZIIZ
-1.87\times 10^{-2} IIIZZI
+2.46\times 10^{-2} IIZZII
-2.81\times 10^{-2} IYIYII\\&
-2.81\times 10^{-2} IXIXII
-2.01\times 10^{-3} IIZIZI
+7.96\times 10^{-4} ZZIIII
+2.81\times 10^{-2} YZYIIZ\\&
+2.81\times 10^{-2} XZXIIZ
+2.16\times 10^{-2} IIYXXY
-2.16\times 10^{-2} IIYYXX
-2.16\times 10^{-2} IIXXYY\\&
+2.16\times 10^{-2} IIXYYX
-2.81\times 10^{-2} YZYZII
-2.81\times 10^{-2} XZXZII
+2.81\times 10^{-2} IYXIXY\\&
+2.81\times 10^{-2} IXXIXX
+2.81\times 10^{-2} IYYIYY
+2.81\times 10^{-2} IXYIYX
+3.83\times 10^{-2} YXIIXY\\&
+2.81\times 10^{-2} IYZYZI
-3.83\times 10^{-2} YYIIXX
-3.83\times 10^{-2} XXIIYY
+2.81\times 10^{-2} IXZXZI\\&
+3.83\times 10^{-2} XYIIYX
+3.83\times 10^{-2} YXXYII
-3.83\times 10^{-2} YYXXII
-3.83\times 10^{-2} XXYYII\\&
+3.83\times 10^{-2} XYYXII
-2.81\times 10^{-2} XIIYYX
+2.81\times 10^{-2} XIIXYY
+2.81\times 10^{-2} YIIYXX\\&
-2.81\times 10^{-2} YIIXXY
\end{aligned}
\end{align}

\end{widetext}

\end{document}